\documentclass[12pt]{spieman}  
\usepackage{amsmath,amsfonts,amssymb}
\usepackage{graphicx}
\usepackage{setspace}
\usepackage{tocloft}

\title{Development of Instruments for Space Exploration Using Meteorological-balloons}

\author[a]{Debashis Bhowmick}
\author[a,b]{Sandip K. Chakrabarti}
\author[a,*]{Ritabrata Sarkar}
\author[a]{Arnab Bhattacharya}
\author[c]{A. R. Rao}
\affil[a]{Indian Centre for Space Physics, 43 Chalantika, Garia Station Rd.,
          Kolkata 700084, India}
\affil[b]{S.N. Bose National Centre for Basic Sciences,
          JD Block, Salt Lake, Kolkata 700097, India}
\affil[c]{Tata Institute of Fundamental Research,
          Homi Bhaba Road, Colaba, Mumbai 400005, India}

\cftpagenumbersoff{figure}
\cftpagenumbersoff{table} 
\begin{document} 
\maketitle

\begin{abstract}
Indian Centre for Space Physics is engaged in studying terrestrial and extra-terrestrial 
high energy phenomena from meteorological balloon borne platforms. A complete payload
system with such balloons is at the most about five kilograms of weight. One has
to adopt innovative and optimal design for various components of the experiment,
so that the data can be procured at decent heights of $\sim 35-42$ km and at the
same time, some scientific goals are achieved. In this paper, we mainly describe
the instruments in detail and present their test and calibration results. We
discuss, how we implemented and tested three major instruments, namely, a
Geiger-M\"uller counter, a single crystal scintillator detector and a phoswich
type scintillator detector for our missions. We also present some flight
data of a few missions to demonstrate the capability of such experiments.
\end{abstract}


{\noindent \footnotesize\textbf{*}Ritabrata Sarkar,  \linkable{ritabrata.s@gmail.com} }

\begin{spacing}{2}   

\section{Introduction}
\label{sec:intro}
Traditionally, for balloon-borne astrophysical observations, large balloons of
several million cubic meters are used (e.g., Ref.~\citenum{yaji09}) with
payloads of several thousand kilograms. These are typically equipped with
ballasts and valves to have long flights of several days to several months
duration. These also typically have, apart from the main instruments, accurate
pointing instruments to acquire data from precise directions. In the other end
of the spectrum, there are meteorological balloons which can carry generally
`use and throw' equipment totaling a few hundred grams for measuring atmospheric
parameters up to a height of $\sim 20-25$ km on a daily basis.

With the advent of modern miniaturized instruments, it is now possible to
explore space using light weight payloads. This aspect has been the major goal
of research by Indian Centre for Space Physics (ICSP) which has systematically
developed a paradigm to study various objects emitting high energy radiation in
space from very light weight meteorological balloons \cite{chak14,chak17}.
Being light weight, these balloons can carry at the most about five kilograms of
payload which must contain not only the main measurement unit, but also the
auxiliary instruments, power-supply, parachutes for re-entry etc. Thus a great
deal of innovation is required to make these low-cost space missions a
success. One of our motivations is to test cubesat and nanosat instruments prior
to actually flying them to space. Being low cost, our procedure is affordable
and is a great learning tool for college and university students.

The instruments in these experiments can be used to measure the intensity of
ionizing radiations, particularly X-rays which is very useful for the study of
Cosmic Rays (CRs), solar activity, X-ray background and accreting compact
objects. It is also possible to detect the high energy Gamma-Ray Bursts (GRBs)
in these kind of experiments. Apart from these extra-terrestrial events,
Terrestrial Gamma-ray Flashes (TGFs) from the cloud formation region of the
atmosphere are other types of interesting and yet to be understood events which
can be recorded by the instruments.

In the present paper, we discuss in details the instrumentation in this
new paradigm of exploring space with balloons of small size and limited
capabilities. As discussed in Ref.~\citenum{chak14,chak17}, the balloons we use
conventionally are rubber weather balloons and often two balloons are tied-up
together to fly a heavier payloads of up to $4$ kilogram reaching a ceiling
altitude of about $35-39$ km. We also use plastic (polyethylene) balloons of
about $7-9$ kg weight which can carry a combined payload of $\sim 6$ kg to
a height of $40-42$ km. We do not use any pointing device and thus we adjust our
launch window to observe the target object(s) for a significant period of time,
unless we want to measure only the CRs. We also tag each photon event (along
with its timing and spectral information) with concurrent attitude of the
payload \cite{chak17,sark19b}. This enables us to compute RA and DEC of the
detector direction during the record of each photon in conjunction with the
instantaneous GPS information of the payload. However, the actual directional
information of the recorded photons are limited by the Field-of-View (FoV) of
the collimator used in the detector which is independent of the detector
direction. Depending on the science goal and experimental conditions, we have
used different collimators with FoV varying between 6-15$^{\circ}$ and sometimes
as wide as 40$^{\circ}$. The measurement of detector or payload direction is
also subjected to instrumental and systematic errors which has been calculated
as $\sim$ 0.3-1.8$^{\circ}$ depending on the rotational speed of the balloon. 
The other major part of the error comes in due to the slewing movement of the
payload and the rate of data sampling for recording. We expect to improve this
in future.

In the next Sec.~\ref{sec:mission} we briefly discuss the experimental
aspects and mission strategies of this novel space exploration program.
Subsequently, in Sec.~\ref{sec:gmc}, \ref{sec:bic} and \ref{sec:phos}, we
describe typical instruments which have been flown. Of course, many of these
flights were dedicated to test the feasibility. For each instrument, we also
present the electronic circuits used, the laboratory tests conducted before
flying and an illustrative flight data of the corresponding detector. Finally,
in Sec.~\ref{sec:conc}, we summarize our results.

\section{A Brief Mission Overview}
\label{sec:mission}
A brief discussion of the experimental strategy with light-weight radiation
detectors has been presented in Ref.~\citenum{chak17}. Study of correlation
between cosmic rays and solar activities using multi-mission data has been
carried out in Ref. ~\citenum{sark17}. Presently, we discuss how each Mission is
executed.

\begin{figure}
\begin{center}
\begin{tabular}{c}
\includegraphics[width=1.0\textwidth]{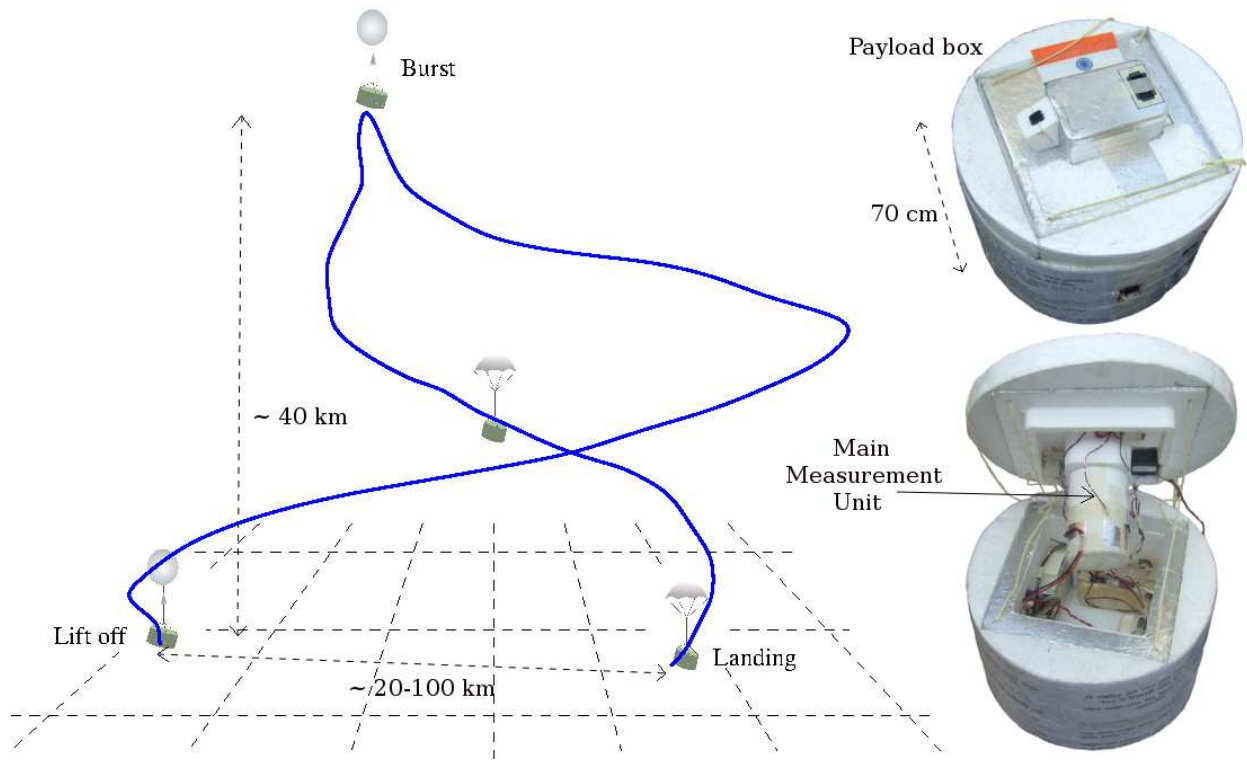}
\end{tabular}
\end{center}
\caption 
{ \label{fig:mission}
(Left:) Schematic drawing of a balloon flight trajectory. (Right:) A typical
payload used in the experiment: external view showing the overall payload
box (top) and internal view showing the main measurement unit and
other ancillary instruments (bottom).}
\end{figure}

As already mentioned in the introduction, the major goal of these
experiments is to measure several extraterrestrial and atmospheric radiation
through light-weight radiation detectors onboard meteorological balloons. The
payload used in this purpose contains the main detector for the radiation
measurement using Geiger-M\"uller (GM) counters or scintillator detectors and
ancillary equipment to supplement the data and help the mission operation. The
carrier is usually one or two (depending on the payload weight) hydrogen filled
rubber balloons or a plastic balloon capable of lifting payloads of $\sim$ 5 kg
or less. The flight generally has no fixed cruising level (no ballasts used) and
goes up to the maximum height till the balloon bursts and comes down with the
help of a parachute (for rubber balloons) or using the torn balloon itself (in
case of plastic balloons). The thermal shielding to the instruments is provided
by using a styrofoam (thermocol) box, which also acts as the payload structure
in which the instruments are embedded. Since unlike a rocket borne
instrument, the frequency of the mechanical vibration during the entire flight
is relatively low, this structure serves quite well, acting as a shock absorber
during the entire flight. At the time of landing, the impact could be a bit
severe. The payload box, along with additional shock absorber system made of
simple hollow plastic cylinders placed at the bottom of the payload, absorbs the
impact efficiently. Since a typical flight lasts for a few hours, the study of
wind pattern is made carefully to ensure that the landing takes place within
about a hundred kilometers of the launch site. Typically, we use two launching
windows: the pre-monsoon window in April-May and the post-monsoon window in
October-November \cite{chak17}. A balloon flight trajectory along with a
typical picture of the payload is shown in Fig.~\ref{fig:mission}.

To ensure cost-effectiveness of our Missions, one of the most important
tasks in this type of missions is to recover the payload on landing. This is
necessary both for retrieving the payload for further use in future missions and
the experimental data which is stored onboard in data storage devices (using
micro-SD cards), as currently we do not transmit the data during flight. The
recovery of the payload relies on our accurate flight path prediction and the
tracking device. The flight path and expected landing location is calculated
ahead of time by giving appropriate weightage to the balloon flight simulator
\cite{habhub} results. We also modify the parachute or balloon lift in order to 
avoid any particular patch (water body, hills, jungles) of land for landing. The
tracking device onboard the payload transmits live location (obtained by the GPS
receiver onboard the payload) to the mobile ground stations on vehicle which
follow the payload near the predicted landing location. As a backup, we also
use an SMS alert system which transmits the payload location on landing to
several payload recovery vehicles.

Since, we cannot afford to place a pointing device due to weight
constraints, a very important issue is the determination of payload attitude.
Apart for the omnidirectional measurements such as atmospheric radiation, the
payload attitude information is crucial to know the incoming direction of the
detected radiation. The attitude measurement instrument is a very light weight
Inertial Measurement Unit (IMU) chip which measure and save the attitude
data at detection of every photon. This data used during the offline data
analysis. The details of the attitude measurement will be published in
Ref.~\citenum{sark19b} (in preparation). However, to have a maximum exposure
time of the source of interest in the detector we need to adjust the launch
schedule and payload tilt angle (payload z-axis w.r.t. zenith) in such a way
that the source approaches as close to the zenith as possible when the payload
is near the maximum altitude. To avoid major corrections due to atmospheric
absorption, we do not observe specific sources which are beyond $\sim
45^{\circ}$ from the zenith.

There is no pressure chamber to protect the detectors from the rarefied
atmosphere at high altitude. However, we conduct extensive tests on the
detectors in simulated pressure and temperature chambers in the laboratory, to
study the effects on them under such extreme conditions. We measure the
atmospheric pressure and temperatures inside the payload and outside using
sensors in each of our flights. These atmospheric parameters up to very high
altitude can be used in long term weather predictions. Additionally, in some of
the experiments an optical sensor (sun-sensor) is implemented to verify if the
sun is inside the FoV of the detector. This brief discussion highlights the key
points of the overall experiment and in the following sections we present the
main radiation measurement units in more detail.

\section{The Geiger-M\"uller Counter}
\label{sec:gmc}
One of the simplest measurements one could do is to measure integrated
radiation counts in the atmosphere. We present the results of miniature
Geiger-M\"uller Counters (GMC) in one of our missions. GMCs have been
traditionally used for such purpose \cite{char75}. Data from several experiments
may be used to study the CR variation with time or location.

\begin{figure}
\begin{center}
\begin{tabular}{c}
\includegraphics[width=1.0\textwidth]{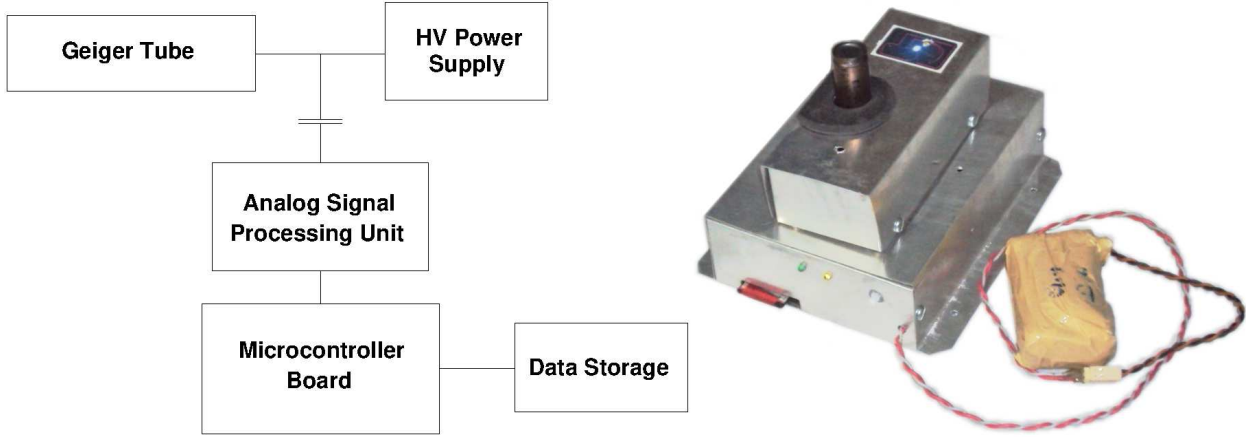}
\end{tabular}
\end{center}
\caption{\label{fig:c2gmblock}
  (Left:) Block diagram of the Geiger M\"uller counter setup and (right:) the
  assembled GM counter with power-supply battery.}
\end{figure}

The count pulses produced in the GM counter, due to the interaction of the
incident $\alpha$, $\beta$ or $\gamma$-rays (particles) are processed and stored
in a micro-SD card. At the same time, we also acquire latitude, longitude,
altitude and GPS time information from the GPS receiver. Hence, when the
payload is launched, we detect high energy radiation counts mainly from the
secondary cosmic rays, as a function of all the three coordinates. The detailed
block diagram of the system is given in Fig.~\ref{fig:c2gmblock}, along with the
picture of the assembled detector featuring the GM tube which is used in the
balloon flight experiment.

We used the GM counter (Model LND712) from \textit{LND, INC.} The detailed
dimensions and specifications of the detector can be found in the data sheet
provided in Ref.~\citenum{lnd}. The dimension of the detector assembly
shown in Fig.~\ref{fig:c2gmblock} is 15$\times$13$\times$14 cm$^3$. The
overall dimension of the total payload box which embeds the detector assembly
and other ancillary instruments is 25$\times$19$\times$17 cm$^3$ and weights
about $1.8$ kg. A single $2.0$ kg category rubber balloon is enough to have the
complete flight of about three hour duration.

\subsection{High Voltage Power Supply}
\label{ssec:gmps}
We generated 500 V required for the GM counter biasing from 5 V DC supply.
The circuit consists of an oscillator followed by a voltage multiplier. We used
a transformer coupling for producing a high voltage and then used a voltage
doubler circuit to achieve our goal.

\subsection{Readout System}
\label{ssec:gmreadout}
The output signal from the GMC anode is taken through a suitable capacitance and
is passed through a resistor-transistor logic circuit to convert the signal
into pulse.

The output of the logical circuit is connected with a microcontroller as an
interrupt signal. It counts the interrupt events per second and stores the raw
format data in a micro-SD card. We use ATmega 32 microcontroller with $\sim$ 11
MHz crystal for the clock. The choice of microcontroller and the clock speed
are sufficient for the radiation count rates we are interested in. A Real Time
Clock (RTC) chip is used to generate an interrupt signal every $1$ s and the
count rate is transferred to the micro-SD card along with the time stamp for record.

\subsection{A sample result of atmospheric radiation counts}
\label{ssec:gmresult}
\begin{figure}
\begin{center}
\begin{tabular}{c}
\includegraphics[width=0.6\textwidth]{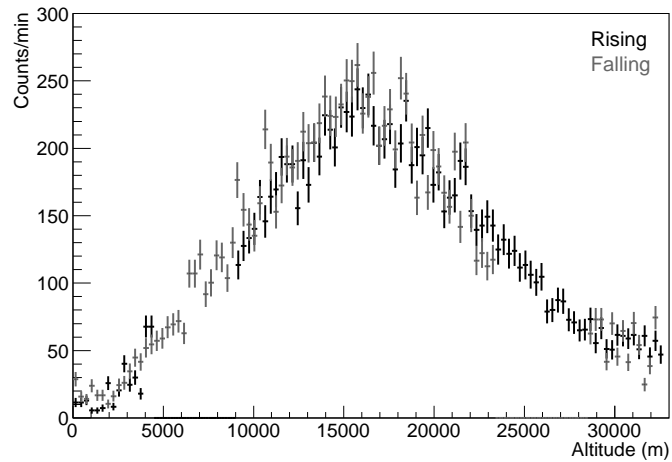}
\end{tabular}
\end{center}
\caption{\label{fig:gmatms}
  Radiation counts detected by GM counter at different heights in the
  atmosphere during the ascend (black) and descend (gray) of the payload.}
\end{figure}

We flew the payload consisting the GM counter as the main measurement unit
on several occasions to measure the integral radiation counts in the atmosphere
at different heights. These atmospheric radiations are mainly due to the
interaction of Galactic cosmic ray particles and solar energetic particles with
the atmospheric nuclei. The window of the GM tube was directed upwards in the
zenith direction and without any collimator. Thus, the detector provides an
omnidirectional measurement of the atmospheric radiation. A result of our
measurement made on 14th May, 2011 (Mission Id. D13) using the GM counter flown
onboard a single rubber balloon is shown in Fig.~\ref{fig:gmatms}, where the
radiation count variation at different heights is plotted. The detected
radiation count rate shows a maximum (Regener-Pfotzer maximum \cite{rege34},
hereafter RP-max) near $\sim$ 16 km during ascent and descent of the detector.
The RP-max arises due to the balance between generation of secondary
radiation from the cosmic ray interaction in the atmosphere and its subsequent
diminution from absorption and decay process at different altitudes. The
radiation count gradually diminishes with height above the RP-max. However, the
atmospheric radiation strongly depends on solar activity, latitude etc. The long
term variability of the atmospheric radiation and its anti-correlation with solar
activity has been studied by us using scintillator detectors in similar experiments
\citenum{sark17}.

\section{Single Crystal Scintillator Detectors}
\label{sec:bic}
For the purpose of X-ray and gamma-ray detection at different energies,
from extraterrestrial sources and in the atmosphere, we use scintillator detectors
mainly with Thallium doped Sodium Iodide (NaI(Tl)) crystal integrated with
a Photo-Multiplier Tube (PMT) for the signal readout. These type of small X-ray
detectors are particularly very useful for the study of solar activity which
emits high intensity  X-rays for which background noise is less severe. We
used integrated detector units (Model 3M3/3 and 2M2/2) with scintillator crystals and
PMTs made by \textit{Saint-Gobain Crystals} \cite{saint}.

\subsection{Detector Specifications}
\label{ssec:bicdet}
In this integrated design, the PMT is optically coupled directly to the
scintillator crystals. The scintillator is mounted in a container (usually
aluminum), and the PMT is shielded with mu-metal. The scintillator container and
mu-metal shield are hermetically sealed together to form a low-mass and light-tight
housing for the detector. The crystal used in 3M3/3 model is cylindrical in
shape and the size is 3'' in diameter and height. The 2M2/2 model contains the
crystal with the same shape except that the size is 2'' in diameter and
height. The weight of the detector including the PMT is $\sim$ 2100 g for 3M3/3
and $\sim$ 1500 g for 2M2/2. The dimension of the assembled detector is about
17$\times$17$\times$40 cm$^3$. The overall payload box has the dimension about
40$\times$40$\times$70 cm$^3$ and weights about 3 kg for 2M2/2 and 4 kg for
3M3/3 model. A double rubber balloon configuration or a single plastic balloon
is enough to have a decent mission with these instruments \citenum{chak17}.

\subsection{Electronics for the Single Crystal Detectors}
\label{ssec:bicelec}
\begin{figure}
\begin{center}
\begin{tabular}{c}
  \includegraphics[width=1.0\textwidth]{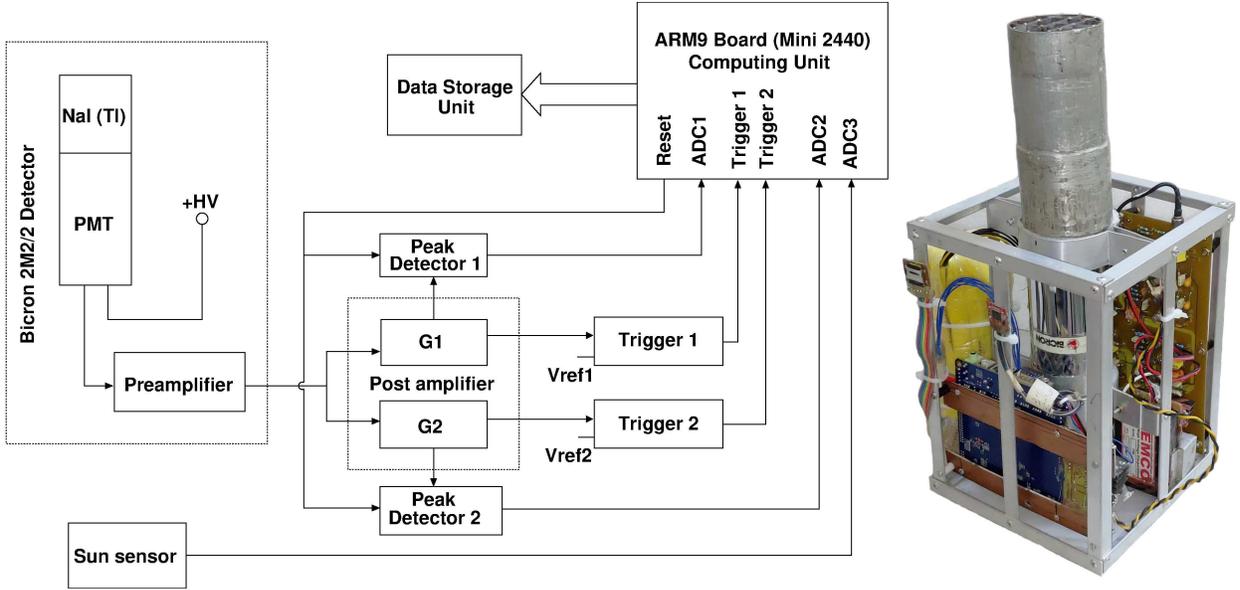}
\end{tabular}
\end{center}
  \caption{\label{fig:b2block}
  The block diagram of the detector and electronic readout system for the
  $2''\times2''$ single crystal scintillator detector. A similar readout system
  is used for the $3''\times3''$ scintillator. The detector assembly along with
  the electronic readout system, collimator and power supply is shown on the
  right.}
\end{figure}

The PMT fitted with the scintillator crystal is provided with a bias voltage
from the high voltage power supply. The signal readout system consists of an
analog front-end circuit, a data processing/acquiring unit, a low voltage
($\pm5$ V and $3.3$ V) DC-DC converter unit and data storage unit. The
overall signal readout scheme for the scintillator detector is shown in
Fig.~\ref{fig:b2block}.

The basic purpose of the electronics signal readout system are:
\begin{itemize}
  \item To generate a high voltage ($\sim 950$ V or $\sim 650$ V for 3M3/3 or
  2M2/2 respectively) for biasing the PMT and a low voltage ($\pm5$ V, $3.3$ V)
  for electronics.
  \item To amplify or process the pulse generated from the detector for the signal
  processing.
  \item To process the pulse signals and record the event data for post-facto
  analysis.
  \item To work in space-like environment in a temperature range namely, 
  $-5^{\circ}$C to $+35^{\circ}$C and qualification in the range from
  $-10^{\circ}$C to $+40^{\circ}$C and near vacuum condition without significant change in
  the performance.
\end{itemize}

The overall electronic circuit may be subdivided into: front-end electronics,
power supply unit, digital signal processing unit and data storage unit which we
discuss in more details subsequently.

\subsubsection{Front-end electronics}
\label{sssec:bicfeelec}
The analog front-end circuit is responsible for processing the analog signal
after getting a pulse signal from the detector. This includes: preamplifier,
post-amplifier, triggering unit and the peak detector.

\paragraph{Preamplifier}
To reproduce the pulses that appear on the anode of the PMT (which are of
short duration and spiky) it is necessary to have a wide band amplifier with
high open loop gain to process fast and low amplitude electrical pulses from
PMT. It is accomplished by using a single high frequency operational amplifier
(op-amp) in inverting amplifier configuration, since the polarity of the
detector’s output is negative in current feedback mode. The operational
characteristics of the preamplifier is summarized in Table~\ref{tab:bicpreamp}.

\begin{table}[ht]
\caption{Specifications of preamplifier.}
\label{tab:bicpreamp}
 \begin{center}
   \begin{tabular}{p{6cm}p{3cm}}
    \hline
    Input supply voltage & $\pm(5\pm0.5)$ V \\
    Rise time of the output pulse & $3$ $\mu$s \\
    Decay time of the output pulse & $10-12$ $\mu$s \\
    Polarity of the output & Unipolar \\
    Voltage gain & $35$ \\
    Saturation level & $5.0$ V \\
    \hline
   \end{tabular}
 \end{center}
\end{table}

\paragraph{Post-amplifier}
The preamplifier output is further amplified in the post-amplifier without
affecting the pulse shape (i.e. decay time). It also provides low impedance to
the following processing and analyzing circuit. Since it is difficult for a
single amplifier to cover a large dynamic energy range of the detector from $15$
keV to $2$ MeV, two different amplifiers with different voltage gains are
provided. First amplifier (G1) covers the lower energy range of $15-140$ keV
while the second amplifier (G2) covers energy range $100$ keV - $2$ MeV. The
operational characteristics of the post-amplifier (G1) is given in
Table~\ref{tab:bicg1}. The G2 amplifier is basically a unity gain amplifier
while the other features are same as G1.

The post-amplifiers have the following features:
\begin{itemize}
  \item The amplifier circuit is provided with $\pm5$ V power supply for its
  operation.
  \item The total power consumption in the front-end electronics is $120$ mW.
  \item The gain and saturation level of the post-amplifier can adjusted
  according to the requirement.
  \item The voltage gain depends only on passive components.
  \item Band pass filters are provided to minimize the low and high frequency
  noise.
\end{itemize}

\begin{table}[ht]
\caption{Specifications of G1 amplifier.}
\label{tab:bicg1}
 \begin{center}
   \begin{tabular}{p{6cm}p{3cm}}
    \hline
    Input supply voltage & $\pm(5\pm0.5)$ V \\
    Rise time of the output pulse & $3$ $\mu$s \\
    Decay time of the output pulse & $10-12$ $\mu$s \\
    Polarity of the output & Unipolar \\
    Voltage gain of two stages & $\sim12$ \\
    Saturation level & $5.0$ V \\
    \hline
   \end{tabular}
 \end{center}
\end{table}

\paragraph{Triggering and peak detector circuits}
The outputs from G1 amplifier is fed to the input of the triggering circuit.
In the present experiment, we use two comparators: one for low ($15-140$ keV)
and another for high ($100-2000$ keV) energy. The low and high energies are
distinguished by two different preset reference voltages while
testing/calibration at the laboratory. Peak detectors are provided for the two
amplifiers (G1 and G2) for analyzing separately.

\subsubsection{Power supply distribution unit}
\label{sssec:bicpowsup}
Power supply distribution unit consists of high voltage and low voltage power
supplies.

\paragraph{High voltage power supply}
The main function of the HV supply is to bias the PMT (at +ve supply). The HV supply
is adjusted such that the PMT gets a bias voltage of $\sim 650$ V for 2M2/2 and
$\sim 950$ V for 3M3/3 detectors. We use EMCO F40 \cite{emco} as the high
voltage module. Since a balloon borne payload reaches $\sim 40-42$ km above
ground, at this near vacuum situation, using such a high voltage requires
potting with a very good quality insulating material to prevent electrostatic
discharge. A silicone elastomeric substrate from Dow Corning \cite{dow} has
been used as a potting material.

\paragraph{Low voltage power supplies}
The low voltage DC-DC power supply unit generates a voltage of
$\pm5$ V and $3.3$ V. The $5$ V is required for the analog front-end while the
on board computing unit works with the $3.3$ V supply.

\subsubsection{Digital data processing, acquisition and control unit.}
\label{sssec:bicarm}
We use a Mini2440 board (ARM9 family processor) \cite{arm9} as the main board
for the data acquisition. The system continuously monitors for the trigger
interrupt signal (event) and when it is found, the computing unit processes the
signal in the following way.

\begin{figure}
\begin{center}
\begin{tabular}{c}
  \includegraphics[width=0.7\textwidth]{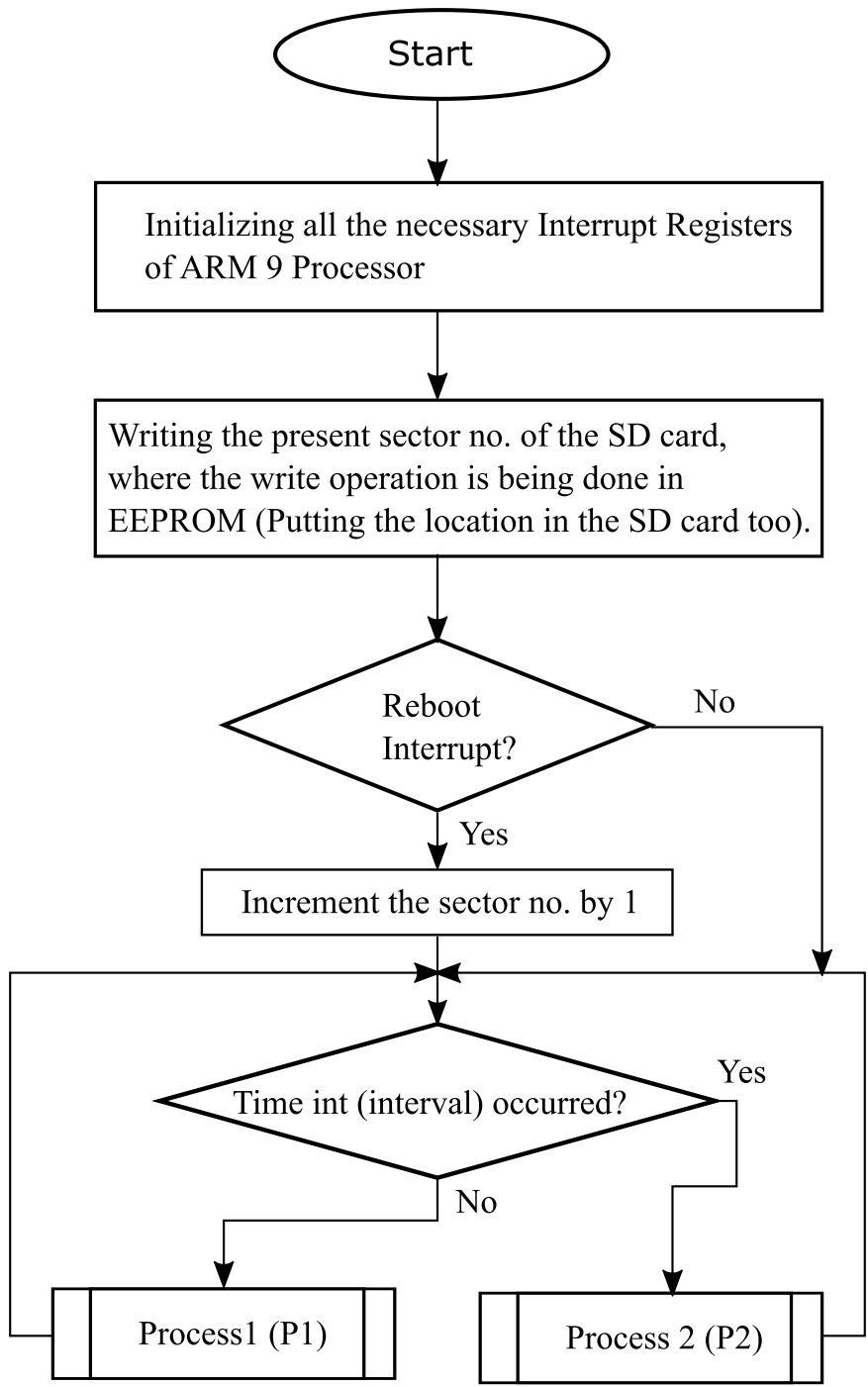}
\end{tabular}
\end{center}
  \caption{\label{fig:armflow}
  Flow diagram for initializing the data processor to process and write
  the events in the SD card. The subsequent acquisition of the event
  (Process1) and writing the data on SD card (Process2) are shown in Fig.
  \ref{fig:armflowp1bic} and \ref{fig:armflowp2}.}
\end{figure}

\begin{figure}
\begin{center}
\begin{tabular}{c}
  \includegraphics[width=0.9\textwidth]{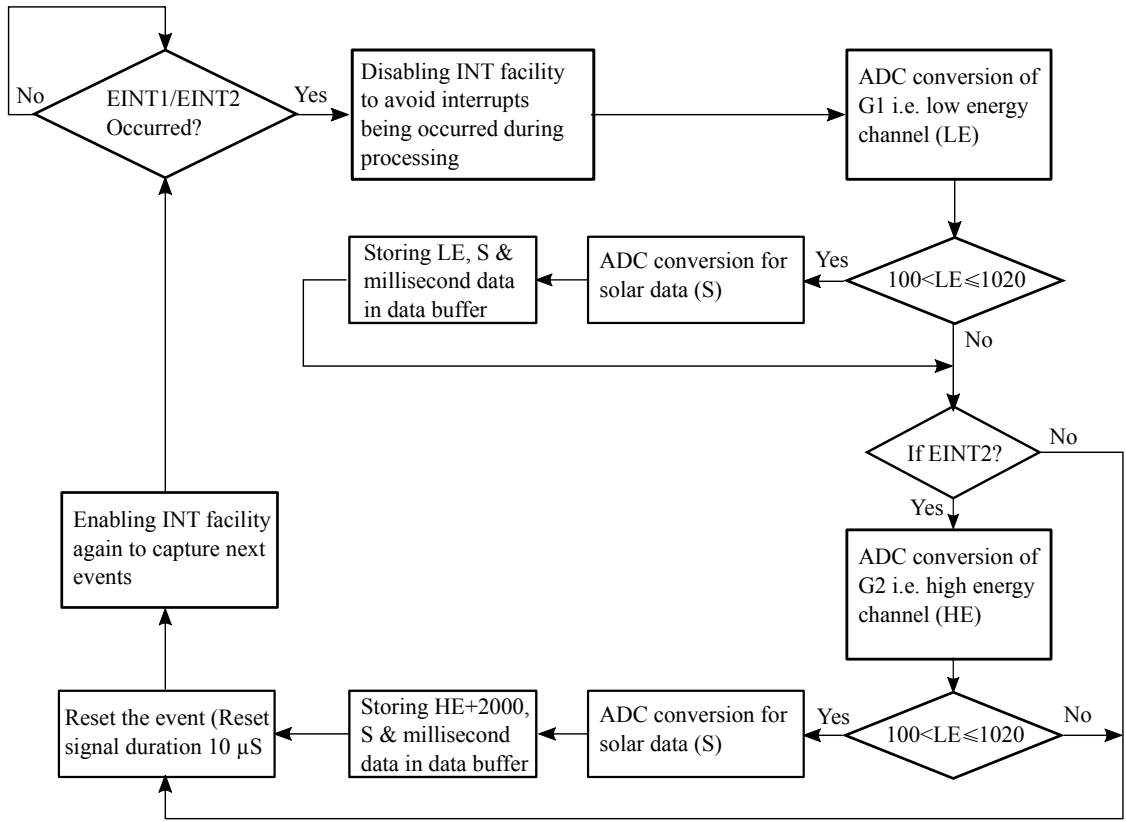}
\end{tabular}
\end{center}
  \caption{\label{fig:armflowp1bic}
  Flow chart for processing (Process1  part in Fig. \ref{fig:armflow}) an
  event both in low and high energy parts and accumulating in the temporary
  buffer.}
\end{figure}

\begin{figure}
\begin{center}
\begin{tabular}{c}
  \includegraphics[width=0.9\textwidth]{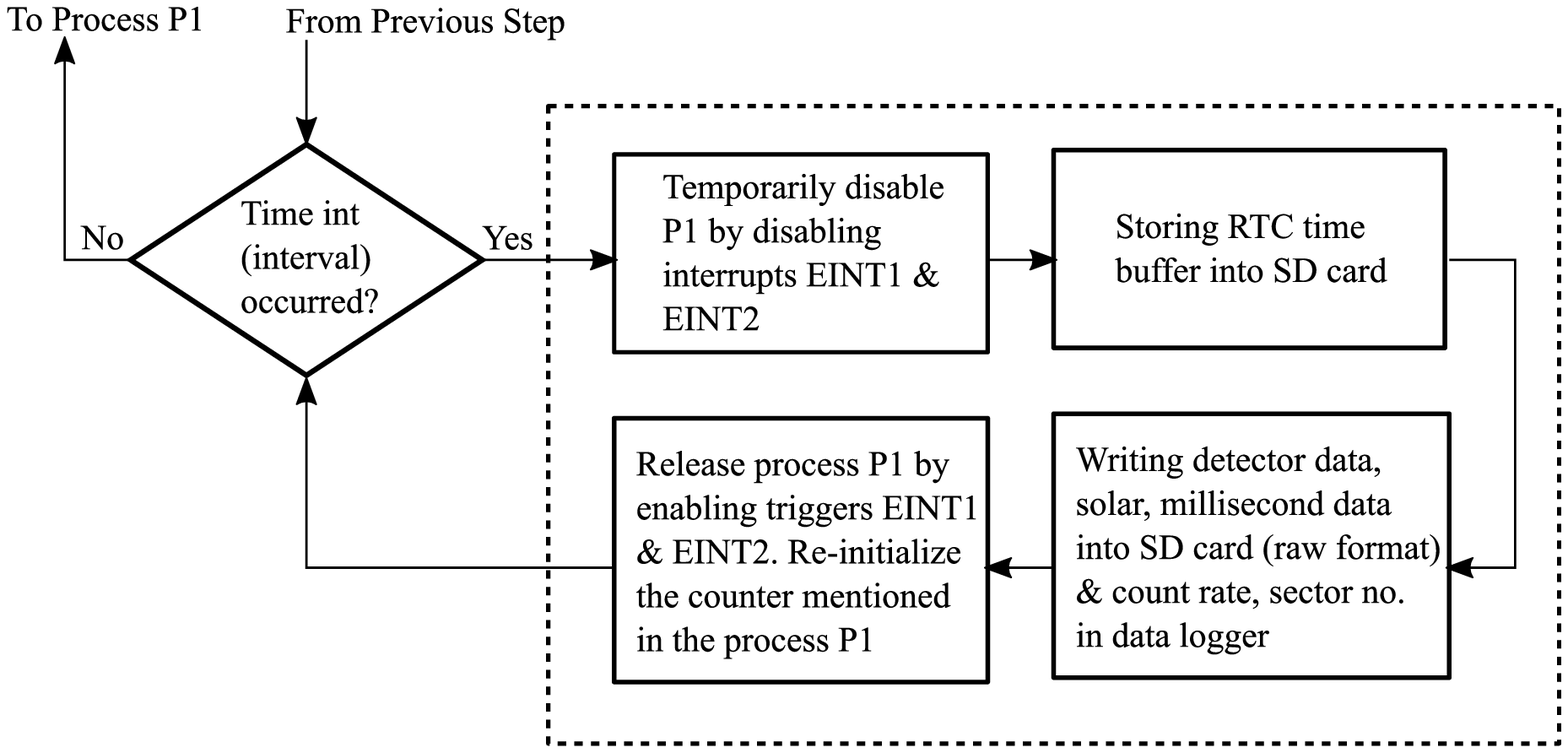}
\end{tabular}
\end{center}
  \caption{\label{fig:armflowp2}
  Flow chart showing the process of writing the buffer data (Process2
  part in Fig. \ref{fig:armflow}) in SD card. The flow input `From Previous
  Step' refers to the input shown in Fig. \ref{fig:armflow}.}
\end{figure}

The system gathers detector Pulse Height (PH) data along with sub-second
time stamp in a temporary buffer. After a pre-defined time interval $\delta t$
all the accumulated events recorded in the buffer are transferred to the on board
micro-SD card and the buffer is cleared. During the first part of $\delta t$, namely
$\delta t_1$, data is procured and temporarily stored in the buffer. During the
second part of $\delta t$, namely $\delta t_2$, the data is transferred to the
micro-SD card. We chose this method, since it is faster to record the data in
the buffer (RAM) rather that directly in the micro-SD card. Depending on the
expected event rate, $dt_1$ and $dt_2$ vary. The whole processing cycle is
shown in Fig.~\ref{fig:armflow}, \ref{fig:armflowp1bic}, \ref{fig:armflowp2}.

Process1 (P1) in Fig.~\ref{fig:armflowp1bic} describes the process of
digitizing each event and storing into the temporary buffer memory (RAM). The
analog PH signals from G1 and G2 are digitized using 10 bit ADC embedded in the
processor (ARM9). However, to exclude the low energy noise information and the
saturated events, we consider only the events in channel 100-1020. This PH
information of the deposited energy along with the time stamp of the
corresponding event and the digitized value from the sun sensor is packed
together and stored the buffer. After storing in the temporary buffer, a reset
pulse of $10$ $\mu$s is issued from the computing unit to reset the event
(discharging the peak-hold circuit at the peak detector) to allow a fresh event
to be captured.

\begin{figure}
\begin{center}
\begin{tabular}{c}
  \includegraphics[width=1.0\textwidth]{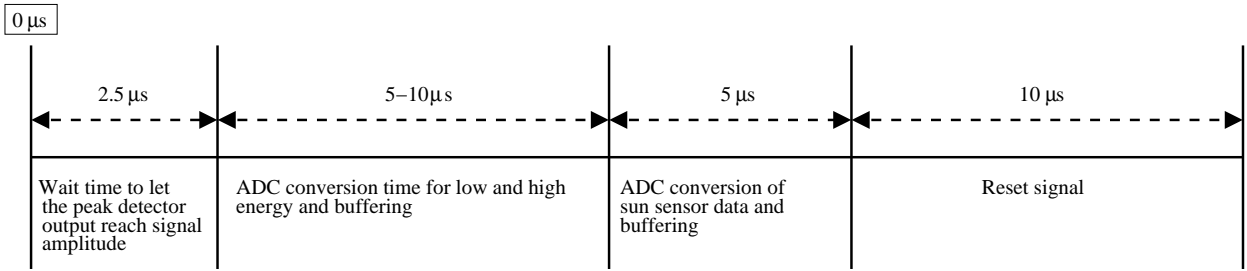}
\end{tabular}
\end{center}
  \caption{\label{fig:bictime}
  Timing diagram of processing an event.}
\end{figure}

The overall time sequence for the processing of a single event is shown in
Fig.~\ref{fig:bictime}.

In Process2 (P2), (flow chart in Fig.~\ref{fig:armflowp2}) the computing
unit writes the content of the temporary buffer memory, acquired during the
designated time interval for Process1, into a permanent storage i.e., an
on-board micro-SD card. After the completion of the data transfer to the micro-SD card
the control is goes back to P1.

\subsubsection{Data format for storing}
\label{sssec:bicform}
\begin{figure}
\begin{center}
\begin{tabular}{c}
  \includegraphics[width=0.7\textwidth]{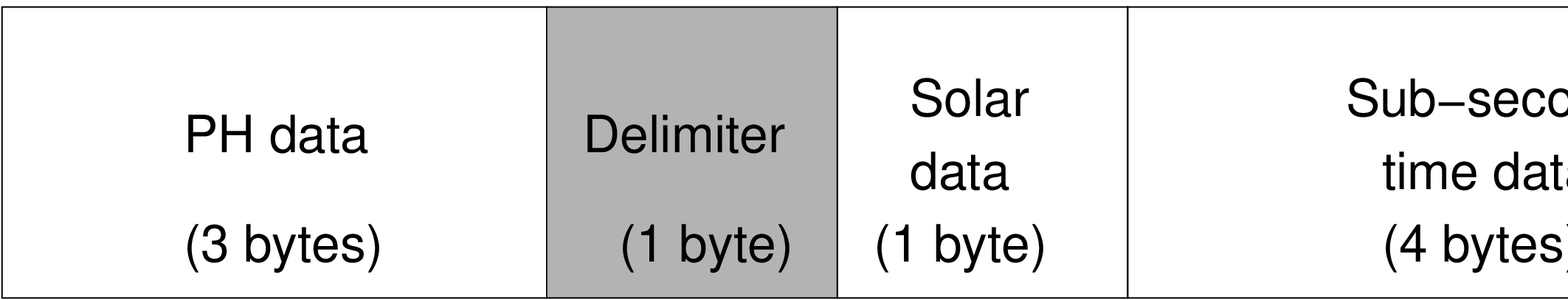}
\end{tabular}
\end{center}
  \caption{\label{fig:bicdataform}
  Showing the packet format for a single event data (total of 10 bytes).}
\end{figure}

Writing to micro SD card requires the computing unit to access the card which is
much more time consuming than accessing the RAM. We therefore optimize
the data structure to reduce the data size so that the required CPU cycles to
write the data in SD card is minimum. To reduce the number of bytes, the
computing unit writes the data in hexadecimal format. The maximum possible
outcome from the ADC is $1023$ (\texttt{0x3FF} in hexadecimal) which requires
$3$ bytes. To distinguish between the low channel energy data (G1) and high channel
data (G2), an offset interval of $2000$ is added to the high channel. Thus the
maximum possible outcome for the high channel becomes $2000$ + $1023$ = $3023$
(\texttt{0xBCF} in hexadecimal). Thus $3$ bytes can be allocated for energy
data. The sun-sensor data from the ADC is binned into $70$ different levels. Thus each
level is $1023$ / $70$ = $14$ (\texttt{0xE} in hexadecimal) requiring only one
byte. The division of a second is done based on the processor clock speed and resolution
of the embedded timer hardware in the computing unit. The maximum value for such
settings is $64859$ which decreases to zero in one second. This requires $4$ bytes. 
One byte is used for a delimiter ‘:’ between energy and sun-sensor data. Finally
a space is used as a delimiter between two events. Thus a total of $10$ bytes 
are required for storing an event in micro SD card. A pictorial representation
of the packet format for a single event is given in Fig.~\ref{fig:bicdataform}.

\subsection{Extracting Readable Data from the System}
\label{ssec:bicdatext}
The  hexadecimal data stored in the micro-SD card is first extracted with a
Matlab \cite{matlab} program to get the data in ASCII format. In the micro-SD card, to save
CPU time cycles and storage space, the bytes are written in the raw format.
The software program running in the ARM9 computing unit keeps track of the
number of memory units that are being written to put the new data in new locations,
thus preventing overwriting.

To extract the raw data stored in the blocks of the card, a low level
access is required. To facilitate the data extraction and for redundancy, another
on-board data storage card is used which stores minimal information of time
stamp, memory address already used in the primary card with the raw data and
the total number of events (low energy, high energy and total counts) in FAT
format. This redundancy allows us to have a quick-look of the data.

\subsection{Laboratory Tests}
\label{ssec:bictest}
Before each Mission, we carry out several laboratory tests on the detector
assembly for its health check and performance. The following standard
functionality tests were conducted on the detector assembly.
\begin{itemize}
  \item Detector performance test under normal laboratory condition.
  \item Energy-channel calibration and resolution of the detector.
  \item Detector performance stability under the low pressure condition.
  \item Detector performance under variable temperature to mimic ascend or descend of the payload.
\end{itemize}
In the following sections, we discuss briefly the tests performed on the detector assembly
before each mission.

\subsubsection{Tests under normal laboratory condition}
\label{sssec:bictest}
The primary check on the detector performance is to test its behaviour under
normal conditions and also tuning its input parameters, such as, HV settings,
reference voltage values in the trigger circuit etc. for its optimum
performance. We used radioactive sources, such as Eu$^{152}$,
Ba$^{133}$, Cs$^{137}$, Am$^{241}$ for calibration.

\begin{figure}
\begin{center}
\begin{tabular}{c}
  \includegraphics[width=1.0\textwidth]{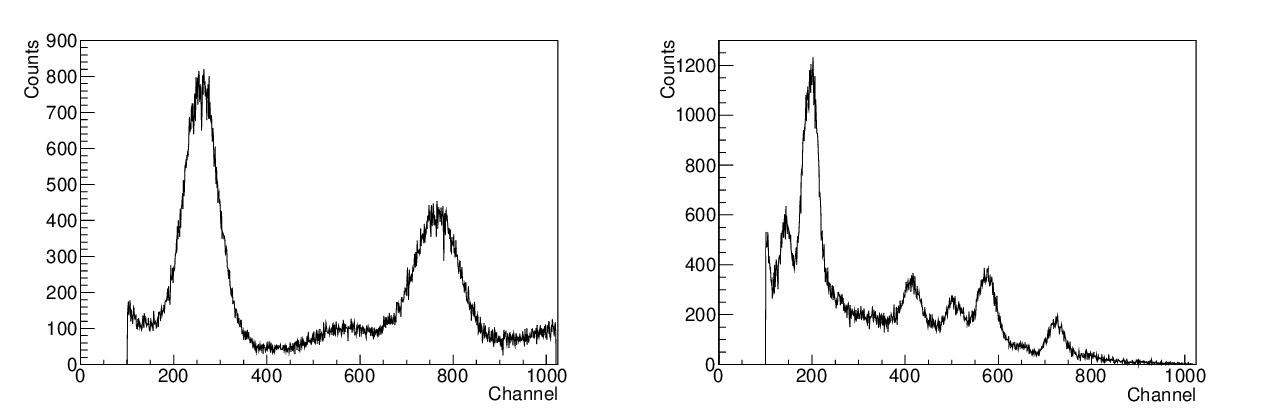}
\end{tabular}
\end{center}
  \caption{\label{fig:biceuch}
  Pulse height spectrum of Eu$^{152}$ source for low (left) and high
  (right) energy events.}
\end{figure}

In Fig.~\ref{fig:biceuch}, we show the pulse height spectrum in ADC channels for
Eu$^{152}$ radiation source, for both low energy and high energy channels.
The applied bias voltage is so chosen that the gain of the system permits
desired energy range within the measurable limit (output $3.3$ V). The gain
parameters can be adjusted from the post amplifier section. An adjustment is
made so that the minimum bias voltage can be obtained with highest resolution
(at 59.5 keV of Am$^{241}$ source). This spectrum shows different radiation
lines, such as, two lines at $39.50$ and $121.9$ keV in the low energy spectrum and
six lines at $245$, $344$, $780$, $960$, $1110$ and $1410$ keV for high energy spectrum.

\subsubsection{Calibration and resolution}
\label{sssec:biccal}
\begin{figure}
\begin{center}
\begin{tabular}{c}
  \includegraphics[width=1.0\textwidth]{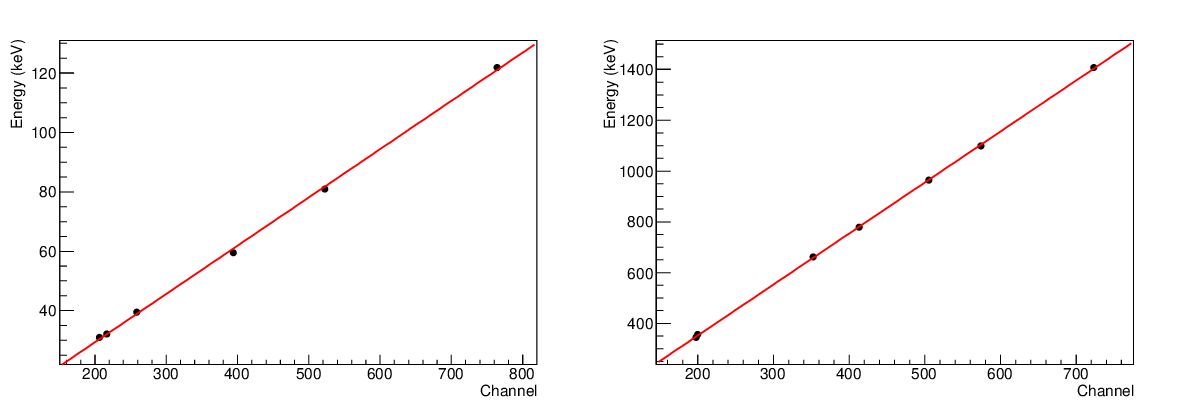}
\end{tabular}
\end{center}
  \caption{\label{fig:biccal}
Low (left) and high (right) energy-channel callibration. The low energy part shows 
a gain factor of $0.16$ keV/channel while the high energy part have the gain 
factor $2.01$ keV/channel.}
\end{figure}

We used different emission lines at known energies from various radio active sources 
to calibrate the detector channels to convert the PH of the events into the
photon energy. This channel-energy relation for low energy events in G1 and high
energy events in G2 are shown in Fig.~\ref{fig:biccal} and they respectively follow
linear relations of the form $E = -3.12 + C * 0.16$ and $E = -49.75 + C * 2.01$.

\begin{figure}
\begin{center}
\begin{tabular}{c}
  \includegraphics[width=0.6\textwidth]{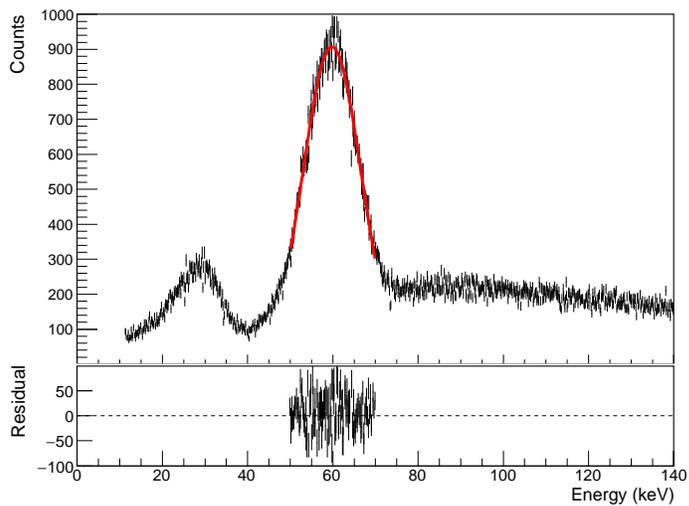}
\end{tabular}
\end{center}
  \caption{\label{fig:bicnaiamen}
  Energy spectrum of the low energy in the detector from Am241
  radiation source. Gaussian fit at $59.5$ keV gives 23.07\% resolution. 
  Residuals to the fit are shown in the lower panel.}
\end{figure}

We calculated the resolution of the detector in a standard way at various
energies by fitting the detected lines using Gaussian functions. The Full Width
at Half Maximum (FWHM) and peak energy (E$_{peak}$) values were obtained from
the fitting. For example, the resolution calculated for the detector at $59.5$ keV
of Am$^{241}$ source is 23.07\% (FWHM/E$_{peak}$). The energy spectrum for this
radiation source along with the Gaussian fit at the $59.5$ keV energy is shown in
Fig.~\ref{fig:bicnaiamen}.

\begin{figure}
\begin{center}
\begin{tabular}{c}
  \includegraphics[width=0.6\textwidth]{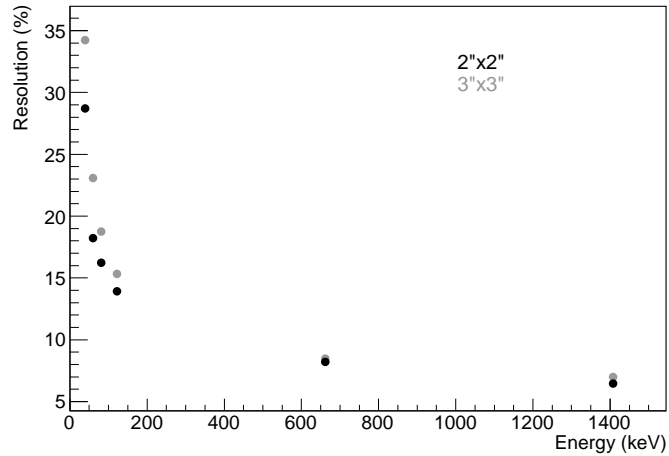}
\end{tabular}
\end{center}
  \caption{\label{fig:bicreso}
  Energy resolution of the $2''$ and $3''$ detector at different energies as
  given in Table~\ref{tab:bicreso}.}
\end{figure}

The resolutions of the $2''$ and $3''$ detectors obtained at various energies
using the radiation source lines are given in Table~\ref{tab:bicreso} and
plotted in Fig.~\ref{fig:bicreso} to show the variation.

\begin{table}[ht]
\caption{Resolutions of the $2''$ and $3''$ detectors at different calibrator line energies.}
\label{tab:bicreso}
 \begin{center}
   \begin{tabular}{ccc}
    \hline
    Energy (keV) & $2''$ resolution (\%) & $3''$ resolution (\%) \\
    \hline
     $39.50$ & $28.72$ & $34.24$ \\
     $59.54$ & $18.22$ & $23.07$ \\
     $80.997$ & $16.23$ & $18.75$ \\
     $121.9$ & $13.91$ & $15.33$ \\
     $661.657$ & $8.20$ & $8.47$ \\
     $1408.0$ & $6.45$ & $6.98$ \\
    \hline
   \end{tabular}
 \end{center}
\end{table}

Stability of gain of the detector for long duration operations was also
tested under normal laboratory conditions for all the detectors. The tests were 
satisfactory. This is explained in more details in Sec.~\ref{sssec:phcal}.

\begin{figure}
\begin{center}
\begin{tabular}{c}
  \includegraphics[width=1.0\textwidth]{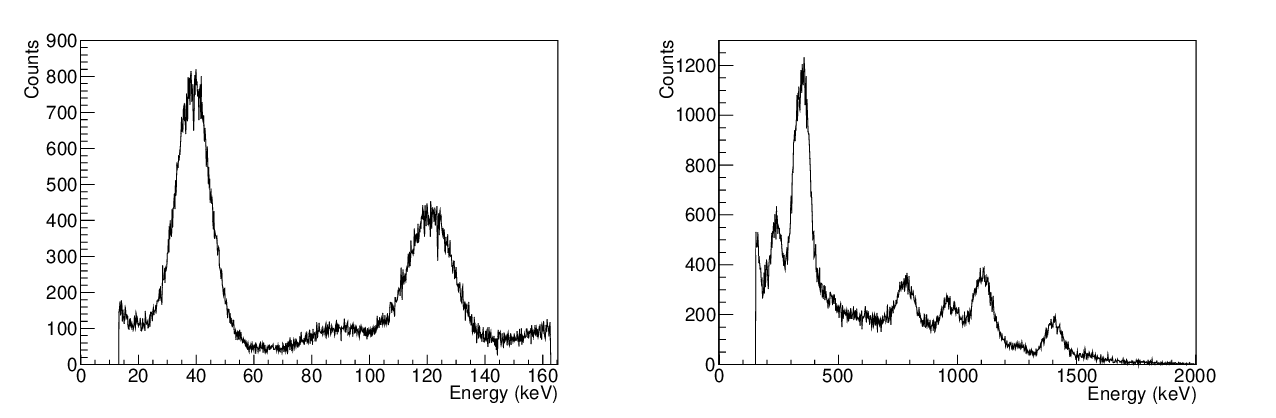}
\end{tabular}
\end{center}
  \caption{\label{fig:bicspeu}
  Measured energy spectrum of Eu$^{152}$ after the calibration. Left and right
  panels are for the low energy (left) and the high energy (right) part.}
\end{figure}

After calibration of the detector, the energy spectrum with Eu$^{152}$ radiation
source is plotted Fig.~\ref{fig:bicspeu}.

\subsubsection{Test under low pressure condition}
\label{sssec:bicvac}
Due to the weight constraint on the payload, we cannot use a pressure
vessel to keep the detector under constant pressure condition. So the detector
is exposed to low pressure condition in the atmosphere up to $\sim 42$ km
during the flight. We performed a test where the detector is kept in a pressure
chamber which reduces the pressure gradually till $\sim 0.5$ mBar was
reached which roughly corresponds to $\sim$ 55 km altitude in the atmosphere.
All the single crystal and phoswich detectors under this test show that the gain
is unaffected by the pressure variation. As an example, the test
result is shown for phoswich detector in Fig.~\ref{fig:vacnaidsp} in
Sec.~\ref{sssec:phvac}.

\subsubsection{Test under temperature variation}
\label{sssec:bictemp}
We conducted tests on the detectors keeping them in a thermal chamber where the
temperature was changed from the room temperature (about $27^{\circ}$C) to $\sim
-10^{\circ}$C inside a low temperature test chamber. Due to the thermal
insulation provided by the thermocol enclosure of our payload, the temperature
inside the payload box is maintained well in this limit during the flight. This
test also shows no significant effect on the detector gain or other parameters
due to progressive changes in temperature.

\subsection{Detection of solar radiation}
\label{ssec:bicflt}
\begin{figure}
\begin{center}
\begin{tabular}{c}
  \includegraphics[width=0.6\textwidth]{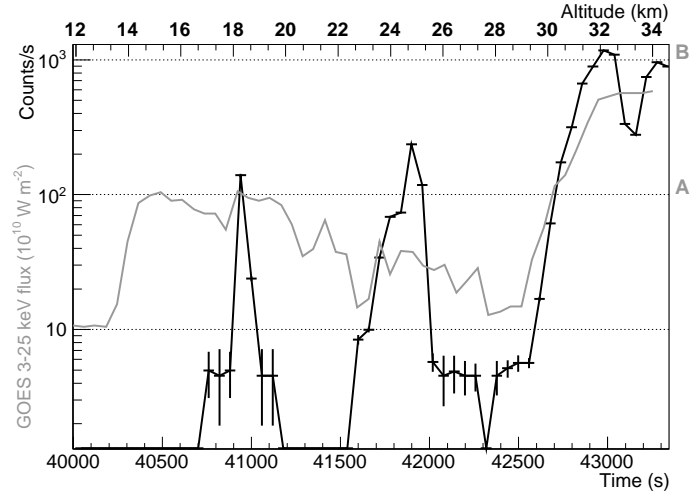}
\end{tabular}
\end{center}
  \caption{\label{fig:bicflt}
  Radiation counts (25-60 keV) (black) in a 2'' single crystal scintillator
  detector onboard a balloon flight during a solar flare and its comparison with
	the GOES data in 3-25keV (gray).}
\end{figure}

We used 2'' and 3'' diameter single crystal scintillator detectors onboard
balloon flights in several missions, for the measurement of atmospheric
radiation due to CR interaction and extraterrestrial radiations. For
example, on 25th April, 2013 (Mission Id. D33), we used a 2'' diameter
scintillator detector onboard a carrier of two rubber balloons tied together, to
measure solar radiation. During this epoch, the sun was in a highly active phase
with frequent solar flares. The experiment was designed and scheduled in such a
way that when the sun is closest to the zenith, the payload is also near its
highest altitude, thereby increasing the solar exposure to the detector and
enhancing the chance of flare detection. Figure~\ref{fig:bicflt} shows the 25-60
keV radiation counts in the detector during a part of the flight. The plot from
the altitude of 12 km till the balloon burst is shown. We also plot the solar
irradiation measured by the detector onboard GOES satellite \cite{swpc} in 3-25
keV range (scaled up by 10$^{10}$), for the sake of comparison. However, our
data is influenced by several experimental and environmental effects which we
state below.

During the time of experiment, the closest approach of sun to the zenith was
$\sim$ 10$^{\circ}$. At the time of the highest payload altitude, the sun was at
$\sim$ 12$^{\circ}$ from zenith. To be able to have the sun inside our FOV
during the experiment we needed to tilt the payload axis by $\sim$ 12$^{\circ}$
w.r.t. the zenith. This optimizes the detector exposure to the sun during higher
payload altitude. However, due to the free rotation of the payload, the relative
position of the sun in detector FoV (40$^{\circ}$) changed (since detector axis
is inclined at 12$^{\circ}$ with the rotation axis). So we expect a variation in
solar exposure and hence intensity of detected solar radiation due to this
rotational motion. This is evident from the $\sim 33km$ altitude data in
Figure~\ref{fig:bicflt} where a deep is seen due to non-detection even when the
flare is on. This type of observations do not affect the spectrum very much,
though corrections of the raw data is needed taking care of the attitude of the
payload. Since the altitude of the payload in the atmosphere is varied with
time, the residual atmosphere causing the attenuation of the solar radiation is
also modified. The light curve of flares depends on the energy of the emitted
photons. The major difference in the light curves detected by GOES and by this
experiment is due to different energy ranges of the detectors. Thus, we see
sharp peaks only when the radiation is of energy higher that the GOES range. We
verified that the spectrum shifts towards lower energy with higher counts, as
the balloon goes up. At lower altitude ($\sim$ 14 to 16 km) we missed a part of
the flare as seen in the GOES data. This is due to the strong  atmospheric
absorption. The detailed analysis of the data will be published elsewhere
\cite{chak19}.

\section{The Phoswich Detector}
\label{sec:phos}
To use an X-ray detector for the study of extra-solar sources where
the intensity of the source radiation is relatively low, it is important to reduce
the background counts in the detector. Only passive shielding for the detector is
not always sufficient and hence phoswich technique is used \cite{pete75} to
reduce the background through anti-coincidence method using two different
scintillator crystals of different pulse decay time. Energy depositions in
different crystals by an event can be identified from the corresponding pulse
shapes so that the partial energy deposition causing significant background in the
primary crystal can be identified and eliminated.

A complete phoswich X-ray detector module includes a phoswich scintillator
detector, an on-board computing unit, power supply, and large data storage 
capability. The entire system weighs only $4.5$ kg. Since it has a very flexible
modular architecture, this payload can be reused after each flight 
and any of its units can be changed/replaced if needed. The dimension of the
assembled detector is about 17$\times$17$\times$31 cm$^3$, while the overall
dimension of the whole payload box is about 40$\times$40$\times$70 cm$^3$ and
weights about 5.8 kg. This payload can be sent only by a $7-9$ kg category
plastic balloon.

\subsection{Detector Specifications}
\label{ssec:phdet}
The heart of the module is a low-energy gamma-ray/hard X-ray detector system.
The detector consists of thallium doped sodium iodide (NaI(Tl)) and
sodium doped cesium iodide (CsI(Na)) scintillator crystals stacked together
and viewed by a PMT. This assembly was produced by M/S Scionix Holland BV, The
Netherlands \cite{scionix}.

The NaI(Tl) crystal is $3$ mm thick and $116$ mm in diameter and the
CsI(Na) crystal is $25$ mm thick with the same diameter. The two crystals are
and the PMT (diameter $76$ mm) is optically coupled to the CsI crystal through
optically coupled and hermetically sealed with an entrance window on NaI side
a light guide of $10$ mm thick. Both the crystals are used in X-ray astronomy as
a special choice of scintillator (with $10^{-2}$ to $10^{-3}$ mole Tl and Na
impurities) by virtue of their following properties:
\begin{enumerate}
  \item Relatively high effective atomic no. ($32$ for NaI and $54$ for CsI
  crystal) and hence is a good absorber of hard X-ray 
  \item Efficient optical light production ($415$ nm wavelength emission from
  NaI and $420$ nm wavelength emission from CsI).
  \item In the current configuration of phoswich detector, the NaI(Tl)
  crystal is sensitive to X-ray photons of $15-100$ keV and the CsI(Na)
  crystal in the energy range of $100-1000$ keV. This is because it resides below the NaI
  crystal which absorbs the photons below 100 keV. Both the crystals are
  sensitive to charged particle background.
\end{enumerate}

The light signal from the CsI(Na) crystal has a different scintillation decay
time ($630$ ns) than that from the NaI(Tl) crystal ($250$ ns) and hence this
distinction may be used to eliminate the events with partial energy deposition
in both crystals. The scintillator signals from both the crystals are used in
anti-coincidence for the background rejection. The high energy gamma-ray and
charged particles which deposit their energy partially in both the crystals are
identified and eliminated by this method.

The interaction of X-ray photons of energy up to $100$ keV with NaI and
CsI crystals is fully dominated by the photo-electric process and absorbed
radiation (secondary electron-hole pair absorbed by the impurities) converts
into light photons (due to the decay of the excited impurities). These 
photons eventually strike the photocathode of PMT (gain $\sim 10^6$) and are
converted into narrow electrical pulse whose magnitude (pulse height) is
proportional to the energy of the incident radiation. The energy resolution
(FWHM) of the scintillator phoswich is expected to be $18\%$ \@ $60$ keV and
Pulse Height (PH) variation across the crystal will be less than $3\%$. The
radioactive isotopes Am$^{241}$ ($59.5$ keV), Eu$^{152}$ ($39.5$ keV, $121.9$
keV and $344.44$ keV), Ba$^{133}$ ($30.97$ keV and $80.997$ keV) and Cs$^{137}$
($32.194$ keV and $661.657$ keV) are used for laboratory calibration of the
detectors.

\subsection{Detector Electronics and Readout System}
\label{ssec:phoselec}
\begin{figure}
\begin{center}
\begin{tabular}{c}
  \includegraphics[width=1.0\textwidth]{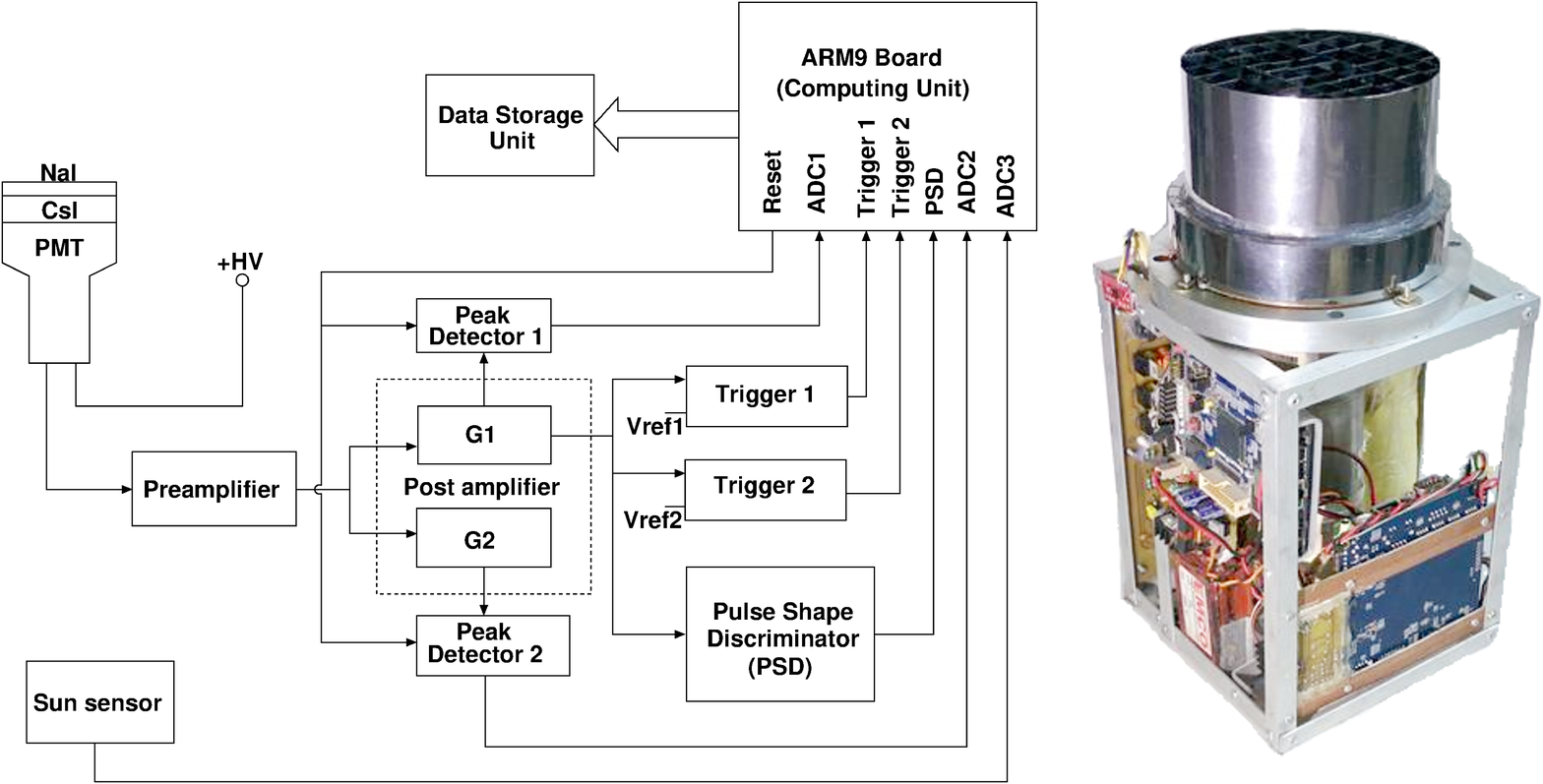}
\end{tabular}
\end{center}
  \caption{\label{fig:phelec}
  (Left:) the schematic block diagram of the electronics and readout system
  for the phoswich detector. (Right:) assembled phoswich detector along with
  the collimator and detector electronics.}
\end{figure}

The schematic block diagram of the overall phoswich detector is given in
Fig.~\ref{fig:phelec} along with a picture of the assembled phoswich
detector. The front-end electronics receives signals from the PMT. These
signals are amplified, digitized and analyzed  in the way discussed below.

\subsubsection{The front-end electronics}
\label{sssec:phfront}
\begin{figure}
\begin{center}
\begin{tabular}{c}
  \includegraphics[width=0.9\textwidth]{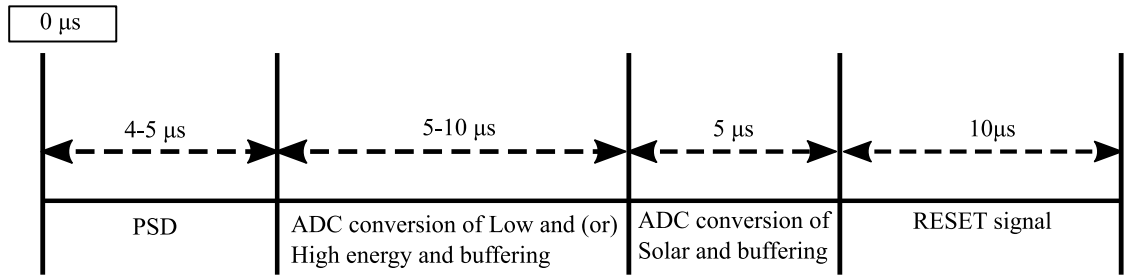}
\end{tabular}
\end{center}
  \caption{\label{fig:pulse}
  Different stages of processing of an event pulse. First $4-5$ $\mu$s
  is used to measure the pulse, next $5$ ($10$) $\mu$s is used to digitize
  (ADC) the pulse height for pulse from single (double) crystal. Another $5$ $\mu$s
  elapses to digitize the sun-sensor data and final $10$ $\mu$s is required to reset the
  system.}
\end{figure}

Signal pulses from the PMT are amplified in a preamplifier and two post
amplifiers (G1 and G2). G1 covers energy range from $15-100$ keV and G2
from $100$ keV up to $1$ MeV. Due to different decay times of pulses
in NaI and CsI crystals, a Pulse Shape Discriminator (PSD) technique is used to
measure the width of the pulses giving the Pulse Shape (PS). An analog
signal at the output of G1 originated from either crystal is used to measure
the pulse shape. The width of the pulse at a certain fractional level of the signal
peak voltage is measured using a counter. The output value of the counter is
recorded as the pulse shape value of the event signal. The pulse height
information is digitized using $10$ bits ADC embedded in the processor providing
1023 channels. The processing of an event in different stages and the
corresponding time lapse is shown in Fig.~\ref{fig:pulse}.

The data is stored into a buffer memory (RAM), for a preset time interval
which depends on the experimental condition with expected maximum count rate.
For example, if the experiment is done under high radiation environment, to
limit the buffered memory, we must have a lower preset time. After that the
buffered data is written on the data storage unit from the memory.

The key features of the phoswich electronics module are:
\begin{itemize}
  \item Amplify phoswich detector output pulses while retaining the original
  shape of the pulse. This is to measure both energy and shape of the pulse to
  allow identification of the origin of the pulse (i.e., NaI or CsI crystal).
  \item Generate regulated power supply from the core power supply ($16$ V,
  $10000$ mAh battery backup) and provide appropriate Low Voltage DC (LVDC) to
  the electronics circuits and High Voltage DC (HVDC) to the PMT of the detector.
  \item Work in space-like environment in a temperature range of $-5^{\circ}$C
  to $+35^{\circ}$C (inside the temperature shielding) and qualify in the range
  from $-10^{\circ}$C to $+40^{\circ}$C without significant change in the
  performance.
  \item Optimize the power and space requirements resulting into less dimension
  and weight so as to be convenient for balloon borne programs.
\end{itemize}

\subsubsection{Digital data processing, acquisition and control unit}
\label{sssec:pharm}
\begin{figure}
\begin{center}
\begin{tabular}{c}
  \includegraphics[width=0.9\textwidth]{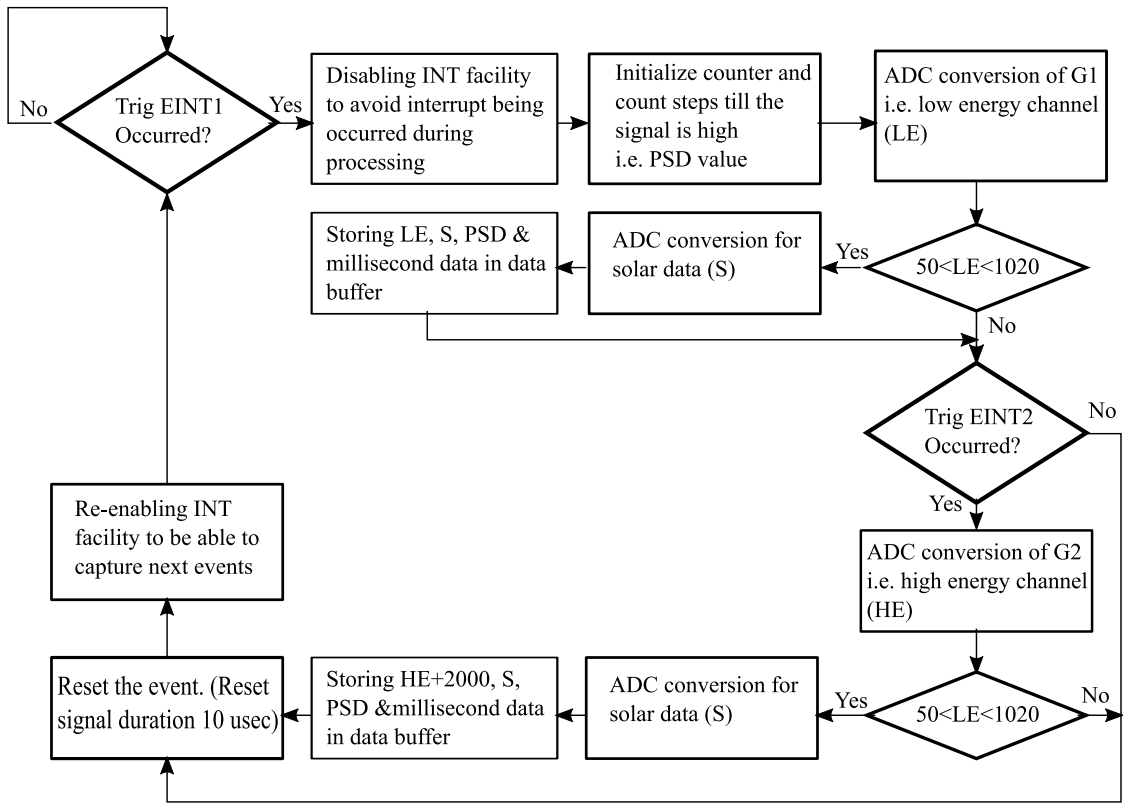}
\end{tabular}
\end{center}
  \caption{\label{fig:armflowp1}
  Flow chart for processing (Process1 part in
  Fig. \ref{fig:armflow}) an event both in low and high energy parts and
  accumulating in the temporary buffer for the phoswich detector.}
\end{figure}

The digital data processing and acquisition system works almost similar to that
described for the single crystal detector in Sec.~\ref{sssec:bicarm}. The only
difference in this case is to generate the PSD value to distinguish the
origin of the events from different crystals from their decay properties. This
difference is shown in Fig.~\ref{fig:armflowp1}. For this case, each event
energy or digitized PH value is also associated with their corresponding PSD
count value along with other components mentioned for the single crystal
detector and packed together as a data packet unit to be stored in the temporary
buffer.

\subsubsection{Data format for storing}
\label{sssec:dataform}
\begin{figure}
\begin{center}
\begin{tabular}{c}
  \includegraphics[width=0.9\textwidth]{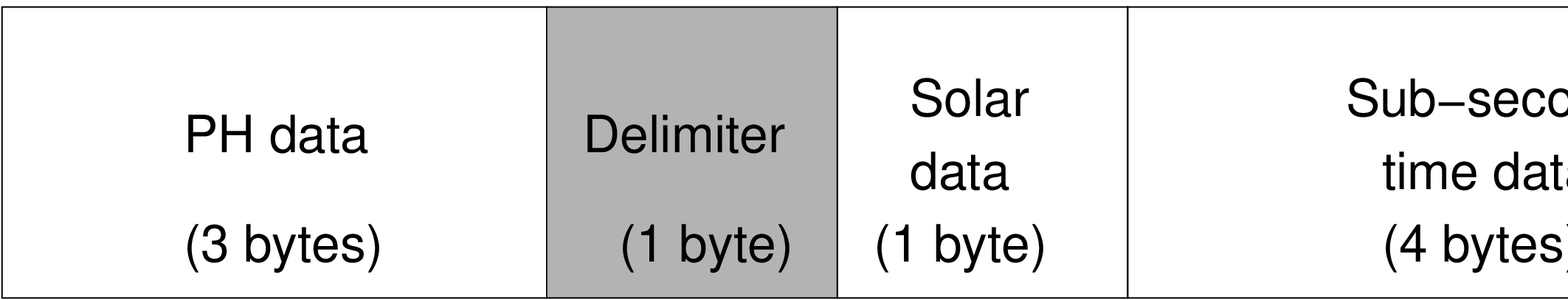}
\end{tabular}
\end{center}
  \caption{\label{fig:dataform}
  The packet format of a single event data in phoswich detector.}
\end{figure}

The format of the event data used to write in the SD card is similar to that
used in single crystal detector described in Sec.~\ref{sssec:bicform}. But in
this case, an extra $2$ bytes for the PSD information and an extra delimiter of
$1$ byte between sub-second and PSD count data are required for each event. So
we need a total of $13$ bytes to record an event data. The division of the data
packet for one event is shown in Fig.~\ref{fig:dataform}.

The extraction of the data from the SD card is similar to that for the single
crystal detector as discussed in detail in Sec.~\ref{ssec:bicdatext}.

\subsubsection{Power supplies for different units (DC-DC converters)}
\label{sssec:phpowsup}
A high voltage DC-DC converter is used for generating bias voltage ($\sim
650$ Volt) for the PMT. A low voltage DC-DC converter is used for the required supply
voltage to drive the front-end electronics of the detector. The whole power
needed by the detector is supplied from the on board  $10000$ mAh, $16$ V battery
power system.

\subsection{Laboratory Tests}
\label{ssec:phlabtest}
We have conducted similar tests in laboratory as discussed in
Sec.~\ref{ssec:bictest} for the phoswich detector, since both types of detectors
are operational under similar conditions. The following functionality tests were
conducted on the detector assembly.
\begin{itemize}
  \item Detector performance test under normal laboratory condition.
  \item Energy-channel calibration and resolution of the detector.
  \item Detector performance stability under low pressure condition.
  \item Detector performance under variable temperature.
\end{itemize}
Here in the following sections we discuss briefly the tests performed on the 
detector assembly.

\subsubsection{Tests under normal laboratory condition}
\label{sssec:phnlc}
\begin{figure}
\begin{center}
\begin{tabular}{c}
  \includegraphics[width=0.6\textwidth]{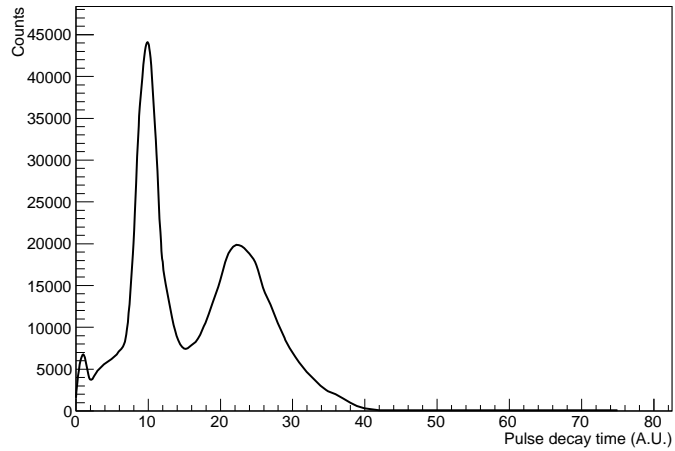}
\end{tabular}
\end{center}
  \caption{\label{fig:psd}
  Pulse decay time plot of the event pulses in phoswich using
  Am$^{241}$ radiation source. The two peaks near $10$ and $24$ are due to
  pulses generated in NaI and CsI respectively and the minimum value near $15$
  is the PS cut value to separate the pulses in two crystals.}
\end{figure}

We used calibrating sources as mentioned in Sec.~\ref{sssec:bictest}. To
discriminate the spectrum from different crystals of phoswich data, first the
PSD data is plotted. The plot helps in distinguishing events coming from NaI(Tl)
or CsI(Na) crystals. Fig.~\ref{fig:psd} shows a PSD plot of the data for the
Am$^{241}$ radiation source. We find two distinguishable peaks in the plot due
to pulses in NaI and CsI with different decay times. The value at the minimum
point near $\sim 15$ between the two peaks is taken as the PS cut value for
separating events from two crystals during the analysis.

\begin{figure}
\begin{center}
\begin{tabular}{c}
  \includegraphics[width=0.6\textwidth]{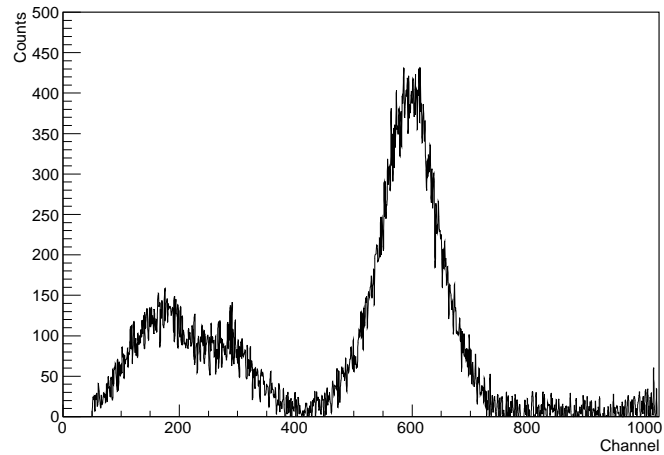}
\end{tabular}
\end{center}
  \caption{\label{fig:naiamchsp}
  Channel spectrum of the events in NaI crystal in low energy from
  Am$^{241}$ radiation source.}
\end{figure}

Fig.~\ref{fig:naiamchsp} shows the channel spectrum of the low energy data
(from G1) for Am$^{241}$ source. The applied bias voltage is so chosen that the
gain of the system permits desired energy range within measurable limit (output
$3.3$ V). The gain parameters can be adjusted from the post amplifier section.
An adjustment is done so that the minimum bias voltage can be obtained with
the highest resolution (at $59.5$ keV of Am$^{241}$ source).

\subsubsection{Calibration and resolution}
\label{sssec:phcal}
\begin{figure}
\begin{center}
\begin{tabular}{c}
  \includegraphics[width=1.0\textwidth]{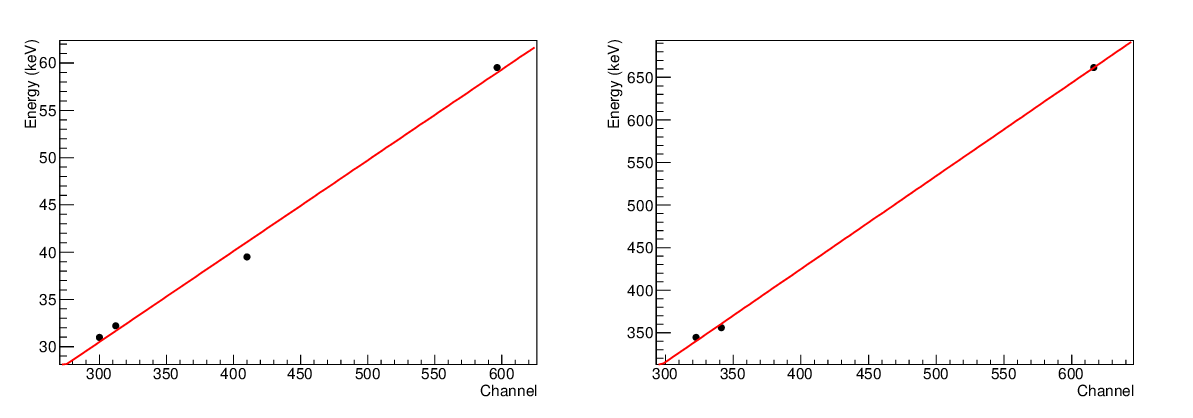}
\end{tabular}
\end{center}
  \caption{\label{fig:phcal}
  Calibration of the detector for low energy events (left) and
  high energy events (right) using different radiation sources. For low energy
  calibration we used lines at 30.97 (Ba$^{133}$), 32.19 (Cs$^{137}$), 39.5
  (Eu$^{152}$) and 59.5 (Am$^{241}$) keV. For high energy calibration we used
  lines at 344.44 (Eu$^{152}$), 356.01 (Ba$^{133}$) and 661.66 (Cs$^{137}$)
  keV.}
\end{figure}

The detector calibration provides the channel-energy relation for low energy
events in NaI (G1) and high energy events (G2) are shown in
Fig.~\ref{fig:phcal}. A linear fitting of the data gives gain factors of
$0.096\pm0.005$ and $1.09\pm0.02$ keV/channel respectively and these
relations are used to convert the PHs into energy deposition information.

\begin{figure}
\begin{center}
\begin{tabular}{c}
  \includegraphics[width=0.6\textwidth]{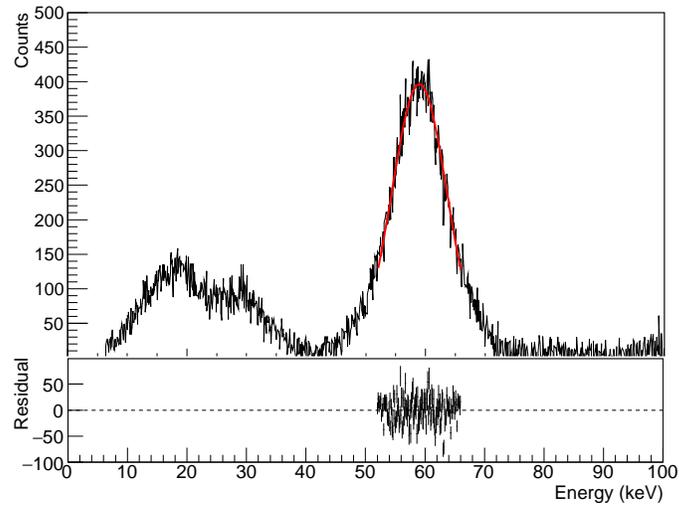}
\end{tabular}
\end{center}
  \caption{\label{fig:naiamensp}
  Energy spectrum of the low energy events in the detector from Am241
  radiation source. The Gaussian fit of the line at 59.5 keV of the low energy
  spectrum gives 18.72\% resolution. The residual plot of the fitting is
  shown in the lower panel.}
\end{figure}

The energy spectrum of the detector for the radiation source Am$^{241}$ using
these calibration information are shown in Fig.~\ref{fig:naiamensp} both for low
and high energy.

We calculated the resolution of the detector at various energies by fitting
the detected lines using Gaussian function in a similar way as discussed in
Sec. \ref{sssec:biccal}. For example, the resolution obtained for the detector
at $59.5$ keV of Am$^{241}$ source is calculated as $18.72\%$. The energy
spectrum in the detector showing the $59.5$ keV line fitted using a Gaussian is
plotted in Fig.~\ref{fig:naiamensp}. These gain factors and resolutions
achieved in our design is comparable to other similar detectors used for
satellite borne space experiments like RT-2 \cite{debn11} and BeppoSAX
\cite{fron97}.

\begin{figure}
\begin{center}
\begin{tabular}{c}
  \includegraphics[width=0.6\textwidth]{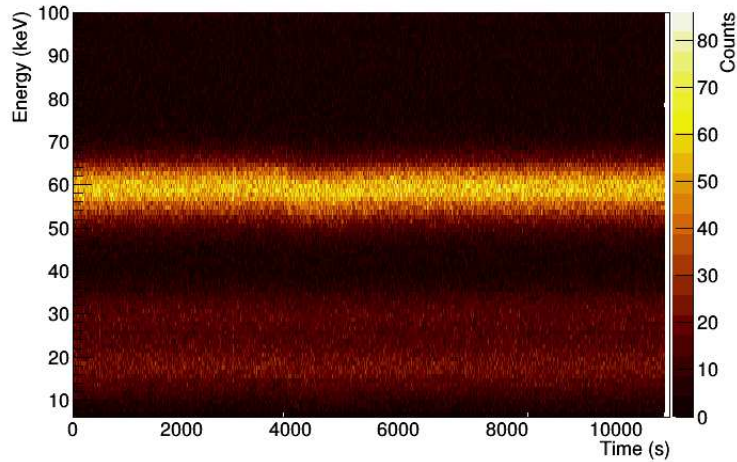}
\end{tabular}
\end{center}
  \caption{\label{fig:naiamdsp}
  Dynamic energy spectrum of the detector
  with time to test the stability of the detector for long operations. 
  Am$^{241}$ source was used.}
\end{figure}

The gain stability of the detector for long duration operation was also
tested under normal laboratory conditions and it was found to be satisfactory. For
example, we show the dynamic energy spectra of the detector with time in
Fig.~\ref{fig:naiamdsp} showing the peaks of Am$^{241}$ radiation source for a
long time of about $11$ ks. The 3$\sigma$ gain variation during this entire
time, as measured w.r.t. the $59.5$ keV line peak, is only about 18.6\% of the
energy resolution at that energy.

\subsubsection{Test under low pressure and low temperature condition}
\label{sssec:phvac}
\begin{figure}
\begin{center}
\begin{tabular}{c}
  \includegraphics[width=0.7\textwidth]{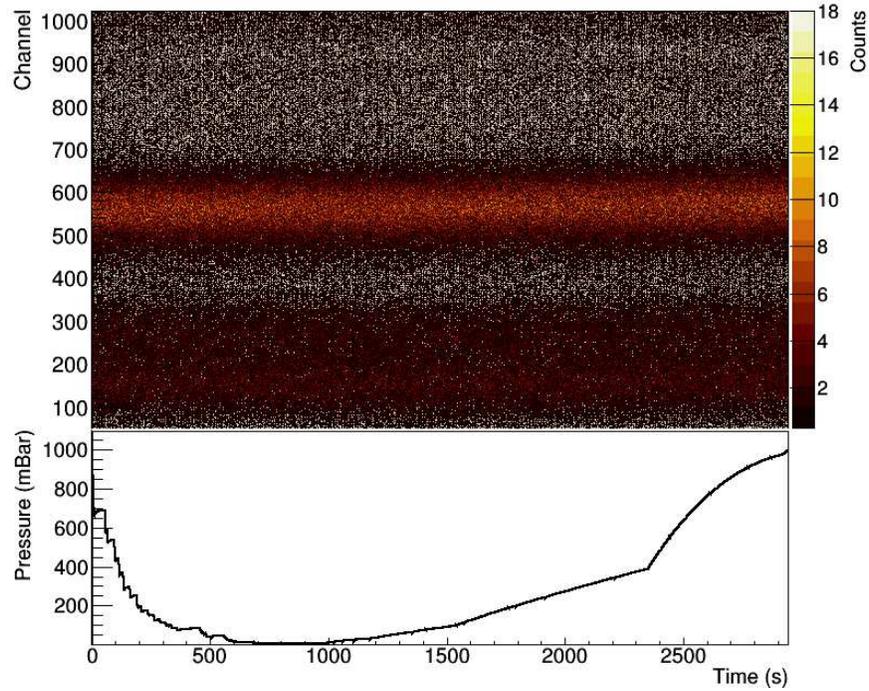}
\end{tabular}
\end{center}
  \caption{\label{fig:vacnaidsp}
  Detector stability test under low pressure condition. The upper
  panel shows the channel spectrum for Am$^{241}$ source with time, while the
  lower panel shows the pressure variation.}
\end{figure}

As in the case of single crystal detector, the phoswich detectors 
were tested under very low pressure condition.
We gradually reduced the pressure inside a pressure vessel containing the
detector till $\sim 0.5$ mBar (equivalent to pressure at $\sim$ 55 km in the
atmosphere) to see any effect of the pressure variation on the detector
operation. Fig.~\ref{fig:vacnaidsp} shows that the detector operation presented
in the upper panel in terms of the dynamic channel spectrum is unaffected by the
pressure changes shown in the lower panel.

The phoswich detector also has been tested under low temperature as the single
crystal detector to show no significant effect on the functionality of the
detector.

\subsection{Detection of Crab radiation by phoswich detector}
\label{ssec:phflt}
\begin{figure}
\begin{center}
\begin{tabular}{c}
  \includegraphics[width=0.7\textwidth]{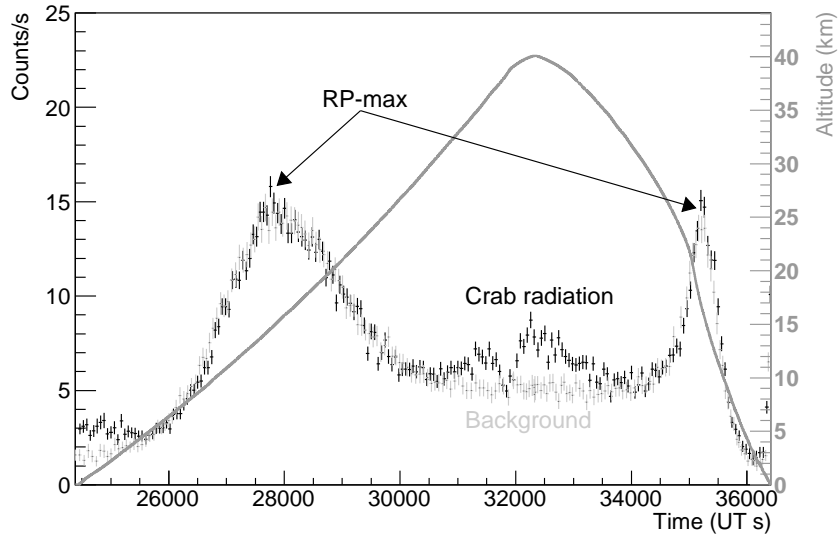}
\end{tabular}
\end{center}
  \caption{\label{fig:phflt}
  Radiation counts (black data points) detected by the phoswich detector
  during its entire flight onboard a plastic meteorological balloon (see text
  for description). Background data in absence of significant astrophysical
  radiation source inside detector FoV is shown by gray data points. The gray
  solid line shows the detector altitude profile.}
\end{figure}

We show, in Fig.~\ref{fig:phflt}, an example of the radiation measurement
using phoswich detector during its flight onboard a plastic meteorological
balloon on 7th May, 2017 (Mission Id. 102). The experiment was designed to
measure radiations from the Crab pulsar. The closest approach of Crab to the
zenith during the time of experiment was about 2$^{\circ}$. So, we scheduled the
experiment in such a way that the payload reaches near its burst altitude when the Crab is
near the zenith. In this way we can minimize the atmospheric absorption of the
source radiation. The detector viewing direction was aligned with the payload
rotation axis (i.e., zenith direction), so there was a small modulation
due to the free rotation of the payload. The sensitivity of the detector is limited by the
atmospheric background radiation and absorption of the source radiation in
the atmosphere which is altitude dependant. From the previous background measurement
experiment, we calculated the minimum sensitivity of the detector at 40 km and
in the energy range of 20-60 keV is $\sim$ 200 mCrab \cite{sark19a}. There was no
other source brighter than this inside the FoV (15$^{\circ}$) of the
detector during the experiment, which was confirmed using other all sky monitor
data onboard satellite (Swift/BAT \cite{krim13}). So, the excess radiation which
is well beyond 3$\sigma$ significance level at the peak from the background is
indeed from Crab pulsar.

The detected radiation count rate in the selected energy range of 25-60 keV
shows the RP-max near $\sim$ 15 km during ascent and descent. The origin of
this radiation in the atmosphere has been discussed in Sec.~\ref{ssec:gmresult}  
and in more detail in Ref.~\citenum{sark17}.
The radiation excess near the highest altitude of the payload indicates the
detection of radiation from the Crab. The sudden dip in the count rate near the maximum
altitude is due to payload attitude change during the balloon burst. The
Figure also shows the background counts in the absence of any significant
astrophysical sources inside the FoV of the detector. This background data was
taken from another mission (Mission Id. D96, on 15th October, 2016) with the same
instrument. The timing information of the background counts were adjusted to
compare with the source data, using payload altitude information of both
missions. A more detailed experimental methodology and results of the temporal
and spectral measurement of the Crab radiation in another similar experiment can
be found in Ref.~\citenum{sark19a}.

\section{Summary}
\label{sec:conc}
Balloon borne space exploration has become an accepted and efficient method
to obtain high energy radiation from space for several decades. With the advent
of miniaturized instruments, it has become possible to send payloads of mass
less than 5-6 kg with the main science data measuring units, location and
attitude measurement units as well as the power supply for the entire mission
and still achieve significant science goals \cite{chak14, chak17, chak19,
sark17}. Though these detectors are modest in size, the technology that is
developed in course of these experiments, can be used to test any new detector
concepts. These detectors can also be used to keep a regular monitoring of
cosmic ray intensity, background etc. apart from the radiation study from
astrophysical objects. In the present paper, we showed how even the normal
payloads such as scintillator detectors or phoswich detectors which could have
been flown to space with regular satellites or large balloons and rockets, may
also be integrated into our low-mass and low-cost space exploration missions. We
do not have pointing systems and do to transmit data to the ground. This led to
innovate new procedures to 'tag' every received photon with its directional
information as obtained from the Inertial Measurement Unit (IMU) chipsets
\cite{sark19b} and especially designed hardware to save data on board in
micro-SD cards. Our present tagging is done with an accuracy of $0.3-1.8$ degree
depending on the rotational speed of the balloon. Our logical circuit
circumvents the need of more sophisticated electronics by simply writing the
data on SD cards in regular time intervals during which data is not collected.
Our data quality is decided by the shielding materials \cite{sark19a}. The test
and evaluation procedure of every instrument is done very strictly. We also
carry out calibration of these instruments in near space conditions by
simulating in-flight pressure and temperature variations. The time-stamped and
attitude tagged photons are analyzed keeping in mind different levels of
atmospheric absorption at different altitudes \cite{sark19a}. These details are
beyond the scope of this paper and will be presented elsewhere.


\subsection*{Disclosures}
The authors have no relevant financial interests in the manuscript and no other
potential conflicts of interest to disclose.

\acknowledgments 
The authors would like to thank Dr. S. Mondal, Mr. S. Chakraborty, Mr. S.
Midya, Mr. H. Roy, Mr. R. C. Das and Mr. U. Sardar for their valuable helps
in various forms during the mission operations and data collection. This work
been done under partial financial support from the Science and Engineering
Research Board (SERB, Department of Science and Technology, Government of India)
project no. EMR/2016/003870. We also thank Ministry of Earth Sciences
(Government of India) for partial financial support. Grant-in-aid from Department of 
Higher Education, Government of West Bengal is acknowledged by DB, SKC, RS and AB to carry out 
the research at ICSP.
 

\bibliography{reference}   
\bibliographystyle{spiejour}   


\end{spacing}
\end{document}